\documentclass[a4paper,11pt]{article}
\usepackage{pos}
\usepackage{tikz-feynman}
\usepackage{tikz, pgf}
\usetikzlibrary{shapes.misc}
\usetikzlibrary{snakes}
\usetikzlibrary{decorations.pathmorphing}
\usetikzlibrary{shapes}
\usetikzlibrary{fit}

\title{IR Constraints on Gyromagnetic Ratios}

\author*[a]{Raffaele Marotta}
\author[b]{Mritunjay Verma}

\affiliation[a]{Istituto Nazionale di Fisica Nucleare (INFN), \\
Sezione di Napoli, Complesso Universitario di Monte S. Angelo ed. 6, via Cintia, 80126, Napoli, Italy. }

\affiliation[b]{ Mathematical Sciences, University of Southampton, \\
Highfield, SO17 1BJ Southampton, UK. }

\emailAdd{raffaele.marotta@na.infn.it}
\emailAdd{m.verma@soton.ac.uk}

\abstract{In Bosonic, Heterotic and Type II theories, by employing soft theorems and string techniques, we analyze the gyromagnetic ratios of massive high spin particles coupled to Abelian gauge fields arising from the compactification of the metric and Kalb-Ramond fields. We derive a universal form for such couplings showing that mixed symmetry states of the leading Regge-trajectory, described by a Young-tableau with two rows, have a gyromagnetic ratio associated with each row.}

\FullConference{%
  Corfu Summer Institute 2021 "School and Workshops on Elementary Particle Physics and Gravity"\\
  29 August - 9 October 2021\\
  Corfu, Greece
}

\begin{document}
\maketitle

\section{Introduction} 

Superstring theories provide  consistent quantum unification of all fundamental interactions including gravity. One of the main features of these models is the presence of an infinite number of massive high spin excitations (HS) whose interactions are consistently described by the softness of the high-energy behaviour of the amplitudes in these theories. In field theory, on the other hand, the study of particles with high spin and their coupling with electromagnetic fields goes back to Fierz and Pauli \cite{FierzPauli:1939} in the late 30' and Belifante \cite{Belifante:1950} in the 50' who proposed the value $ g=1/s $ for the gyromagnetic ratio of  particles with spin $s$. This value is consistent with the minimal coupling substitution but disagrees with the experimental value of the g-factor of the W-bosons and from the high-energy behaviour of the Compton scattering which suggests the value $g=2$ for all the spinning particles \cite{Weinberg:1970}. The disagreement is cured by  adding non-minimal couplings
 among high spins states and the field strength of the gauge field.
The coefficient of the first non-minimal coupling 
in the derivative expansion of the effective action determines the gyromagnetic ratios of the HS-states.
The value $g=2$ also emerges in many other different contexts in both string \cite{Ademollo:1974,ArgyresNappi:1989, Ferrara&Porrati&Telegdi,9809142} and field theories \cite{BargmannMichelTelegdi:1959,Holstein1,Holstein2, 1204.1064}. Moreover, it turns out to be the only consistent value possible with the propagation on constant electromagnetic backgrounds \cite{Cortese:2013lda,Rahman:2016tqc,Porrati:2009bs,Porrati:2010hm,Kulaxizi:2012xp} . There are some exceptions to this natural value. These arise for HS-states coupled with gauge fields resulting from the Kaluza-Klein (KK) compactification of higher-dimensional gravity or supergravity theories\cite{HIOY:84,9801072}  as well as string theories defined in different backgrounds \cite{9502038,9508068,0104238,Sen:1994eb,9612015}.

In these notes motivated by these results, we analyse gyromagnetic factors for massive spinning excitations charged with respect to Abelian gauge fields arising from the compactification of the graviton and anti-symmetric field. Two completely uncorrelated approaches are followed to obtain the gyromagnetic factors, namely, soft theorems and explicit string calculations in Bosonic, Heterotic and Type II theories.

Soft theorems are universal relations between amplitudes with and without massless particles carrying low or soft momentum \cite{SAGEX,Weinberg:1965,Strominger,2203.07957}. They assert that amplitudes with a massless particle having low  momentum are obtained by acting,  with suitable operators,  on amplitudes with only hard or finite energy states. Recently, from different perspectives such as asymptotic symmetries \cite{Strominger,2011.04420}, gauge \cite{1406.6987,2005.05887} and diffeomorphism \cite{1706.00759,1707.07883,1707.06803} invariance, it has been shown that the soft graviton and photon behaviour are universal at leading order in momentum expansion. The universality extends to subleading order only for gravitons in spacetime dimensions larger than four. In d=4 dimensions, the presence of infrared divergences in loop amplitudes modifies the soft graviton behaviour adding logarithmic corrections to the soft factors \cite{1804.09193}. Lower-order soft graviton and gluon operators depend on the angular momentum and spin operators. Therefore, it is not surprising that one can read the gyromagnetic ratios of the finite energy states from the subleading soft expansions. Thus, these restrictions would arise from the IR behaviour of the theory as opposed to the high-energy Compton scattering proposed by Weinberg to restrict the gyromagnetic ratios. In other words, the IR constraints also put restrictions on the gyromagnetic ratios which are consistent with the results coming from the UV constraints. 
 
We consider the subleading soft graviton theorem in arbitrary $d+1$ spacetime dimensions, with $d\geq 4$, and with only gravitons as  finite energy states. We compactify the theory on a one-dimensional circle.  Compactifying the higher dimensional graviton yields a lower dimensional graviton, a massless vector, a scalar, and an infinite tower of massive KK spin-2 states. From the $d+1$ dimensional soft graviton theorem we, therefore, obtain the vector and scalar soft theorems in $d$-dimensions, with massless and massive spin-2 particles as hard states. The case $d=11$ is special because the  resulting compactified ten-dimensional theory is the Type IIA superstring  theory with the vector being identified with the Ramond-Ramond  $1$-form field, the scalar being identified with the dilaton and the KK-massive modes of the metric with bound states of D0 branes\cite{Witten:1995ex,sen_m_theory}. In this case, we generalise the soft theorem  of the dilaton with hard states carrying KK-charges, and we obtain a new soft theorem for the RR 1-form. These new results are also confirmed by explicit amplitude calculations performed in the $d$-dimensional theory obtained by compactifying   the $d+1$ Einstein-Hilbert action and  keeping the entire tower of massive KK-excitations. The KK spin-$2$ modes are described by the Fiertz-Pauli action, whose interaction with the Abelian vectors is introduced by the minimal coupling procedure\cite{Fierz&Pauli,cho&zho,Deser&Waldron}.  As mentioned above, this procedure is ambiguous and requires the presence of an unfixed gyromagnetic coupling between the vector and the massive {spin-$2$} fields. This arbitrary coupling is determined by comparing explicit amplitude calculations, involving a vector interacting with an arbitrary number of massive spin-2 particles, with the corresponding results obtained by compactifying the soft theorem. The two approaches are consistent only with the value g=1 of the gyromagnetic ratios of the massive KK-states. This result confirms previous calculations made in different contexts\cite{HIOY:84,9801072,9405117}.

The above discussion was for the supergravity theories. However, we can also consider string theory which describes a consistent theory of quantum gravity and is formulated in the critical space-time dimensions. The extra dimensions are compactified to define the theories in the observed
$d=4$ space-time dimensions. The compactification introduces new gauge fields arising from massless higher dimensional tensors with one index along the non-compact directions. We shall consider the $U(1)$ gauge fields emerging from the toroidal compactification  of the graviton and Kalb-Ramond fields. Massive string states having winding and KK charges, 
 couple minimally and non-minimally with these fields. These interaction terms can be determined either from the string 
Hamiltonian compactified on a generic toroidal background 
or from the momentum expansion of compactified three-point amplitude involving an NS-NS massless field and two massive spinning particles.
 Both approaches have been followed and a universal formula for the gyromagnetic ratio has been derived in Bosonic, Heterotic and Type II 
theories  under toroidal compactification. It turns out that states with only KK or winding charges have gyromagnetic ratios $g=1$. 
Mixed symmetry HS-states laying on the first Regge-trajectory and described by Young-Tableau diagrams with two rows, 
 display two gyromagnetic couplings (one for each row) which are completely determined by their left and right spins and momenta.

The results discussed in this proceeding are extensively analysed  in \cite{ 1911.05099,2102.13180}  with all the details on their derivation.

The rest of the draft is organized as follows. In section \ref{notation}, we introduce our conventions for the gyromagnetic factor and some useful notation for performing the string calculations. In section \ref{soft}, we consider the compactification of d+1 dimensional soft graviton theorem and derive the lower dimensional 
 vector soft behaviour which is needed to discuss the gyromagnetic ratios. In section \ref{gyrofromstring}, we describe the gyromagnetic ratios of massive particles which arise in toroidal compactification of string theories. We end with conclusion in section \ref{Conclusion}.

\section{ Gyromagnetic Ratios  and Symbols in String Theory}
\label{notation}
The gyromagnetic ratios are related to the  couplings that appear in the effective actions giving the interaction between massive spinning particles and $U(1)$ gauge fields. To lowest order in the derivative expansion, these couplings are described by the action
\begin{align}\label{APhiPhi}
    \mathcal{V}_{A\Phi\Phi}\sim iqA_\mu\left[\Phi^*\cdot(\partial^{\mu}\Phi)-(\partial^\mu\Phi^*)\cdot\Phi\right]+\frac{i\alpha}{2} F_{\mu\nu}(\Phi^{\mu}\cdot\Phi^{\nu})\,,
\end{align}
where the ``$\,\cdot\,$'' denotes the contraction among  the indices  of the fields. The gyromagnetic ratios can be obtained from  the ratio of the coefficient $\alpha$  that appears in front of the  first non-minimal term  and the charge appearing in the minimal coupling
\begin{align}
    g\sim\left|\frac{\alpha}{q}\right|\,.
\end{align}
In $4$-dimensions all the HS-states are classified as totally symmetric fields and they can be conveniently represented using the generating function notation $\Phi(u)=\tfrac{1}{s!}\Phi_{\mu(s)}u^{\mu(s)}$. The generic non minimal coupling is then  written in the form
 \begin{align}\label{gyroSymm}
    \mathcal{V}_{A\Phi\Phi}^{N.M.}=\frac{i\alpha}{4}\, F_{\mu\nu}\left\langle\Phi|S^{\mu\nu}|\Phi\right\rangle\,,
\end{align}
with the inner product given by
\begin{align}
    \left\langle\Phi_1|\Phi_2\right\rangle=\exp\left(\partial_{u_1}\cdot\partial_{u_2}\right)\ \Phi_1(u_1)\Phi_2(u_2)\Big|_{u_i=0}\,.\label{2.6}
\end{align}
Here,  we have introduced the totally symmetric product of $u$'s as
 $u^{\mu(s)}=u^{\mu_1}\cdots u^{\mu_s}$ and the spin operator with $ 
 S^{\mu\nu}_u=u^\mu\partial_{u}^\nu-u^\nu\partial_{u}^\mu$.
  The generalization of the coupling to a generic mixed symmetry state is now straightforward; a generic mixed-symmetry state as a generating function of auxiliar variables $u_i$, turns out to be
 \begin{align}
    \Phi(u_i)=\frac1{s_1!\cdots s_n!}\phi_{\mu_1(s_1)\mu_2(s_2)\ldots\mu_n(s_n)}u_1^{\mu_1(s_1)}\cdots u_n^{\mu_1(s_n)}\,,
\end{align}
supplemented by the the irreducibility conditions $u_{i}\cdot\partial_{u_{i+k}}\Phi=0$ for all $i$ and $k>0$.  For each auxiliary variable, it is convenient  to define a spin operator
\begin{eqnarray}
 S_{u_i}^{\mu\nu}=u_i^{\mu}\partial_{u_i}^{\nu}-u_i^{\nu}\partial_{u_i}^{\mu}\, .\label{5}
 \end{eqnarray}  
 Using the above notation, the natural generalization of the gyromagnetic couplings involving  mixed symmetry states can be written as
\begin{align}\label{standardGyro}
    \mathcal{V}_{A\Phi\Phi}=\frac{i}4\alpha\,F_{\mu\nu}\Big\langle\Phi\Big|\underbrace{\sum_{j=1}^nS_{j}^{\mu\nu}}_{S^{\mu\nu}}\Big|\Phi\Big\rangle\, ,
\end{align}
with $S^{\mu\nu}$ being the total spin. This formalism is easily translated to represent HS-states in string theory. In Bosonic string, for example, a generic closed string state is the factorized product of the left and right sectors according to
 \begin{align}
|\phi\rangle&= {\cal N}_0\ \phi_{\mu_1(s_1) \dots \mu_p(s_p)\bar{\mu}_1(\bar{s}_1)\dots  \bar{\mu}_q (\bar{s}_q)}\, \alpha^{\mu_1(s_1)}_{-n_1} \dots\alpha^{\mu_p(s_p)}_{-n_{p}}\,\bar{\alpha}^{\bar{\mu}_1(\bar{s}_1)}_{-\bar{n}_1}\dots\bar{\alpha}^{\bar{\mu}_{q}(\bar{s}_q)}_{-\bar{n}_{q}}\, |0,\,\bar{0},\,p\rangle
\label{string_states}
\end{align}
with  $\alpha$ being the string oscillators acting on the ground state of the theory. By introducing auxiliary commuting variables $w_i$,  known as  symbols, in  analogy with the generating function notation, we  represent the generic closed string state in the form
\begin{align}\label{mapping}
   |\phi\rangle\to \frac1{s_1!\cdots s_p! \bar{s}_1!\cdots \bar{s}_p!}\,\phi_{\mu_1(s_1) \dots \mu_p(s_p)\bar{\mu}_1(\bar{s}_1)\dots  \bar{\mu}_q (\bar{s}_q)}\, w^{\mu_1(s_1)}_{n_1} \dots w^{\mu_p(s_p)}_{n_{p}}\,\bar{w}^{\bar{\mu}_1(\bar{s}_1)}_{\bar{n}_1}\dots\bar{w}^{\bar{\mu}_{q}(\bar{s}_q)}_{\bar{n}_{q}}\,.
\end{align}
A generic mixed symmetric is then written in the compact notation as follows
\begin{align}\label{basis}
    \frac1{s_1!\cdots s_p! \bar{s}_1!\cdots \bar{s}_p!}\, (u_{n_1}\cdot w_{n_1})^{s_1} \dots (u_{n_p}\cdot w_{n_p})^{s_p}\,(\bar{u}_{\bar{n}_1}\cdot\bar{w}_{\bar{n}_1})^{\bar{s}_1}\dots(\bar{u}_{\bar{n}_q}\cdot\bar{w}_{\bar{n}_q})^{\bar{s}_q}\, ,
\end{align}
where we have decomposed the string polarization tensor in the  product of the left and right components and represented  it in terms of the  auxiliary variables $u_i$ , i.e. $\phi_{\mu_1(s_1) \dots \mu_p(s_p)}= u_1^{\mu_1(s_1)}\cdots u_n^{\mu_1(s_n)}$ with a similar expression for the right sector. In these notes, we will use this convenient representation of the string states which is easily extended to the Fock-space of the physical states of the  Heterotic and closed superstring theory.

\section{Compactification of Soft Graviton Theorem}
\label{soft}

Gravitational soft theorems are relations between $M_{n+1}$ amplitudes,  involving a graviton with small momentum $q$ and $n$ arbitrary particles with finite momentum $p_i\ (i=1,\cdots,n)$,  and $M_{n}$-amplitudes involving only finite energy particles. The $n+1$ amplitudes, in this infrared regime, are obtained by acting with soft operators $\hat{S}^{m}$, $m=-1,0,1$, on amplitudes without the soft particle 
\begin{eqnarray}
\mathcal{M}_{n+1}(q;\{p_i\})
\ =\ \kappa_{d+1}\left[\hat{S}^{(-1)}+\hat{S}^{(0)}+\hat{S}^{(1)}\right] \mathcal{M}_{n}(\{p_i\})+O(q^2)\, .\label{1soft}
\end{eqnarray}
Here $\kappa_{d+1}$ is the $d+1$\footnote{Here, we consider the case $d\geq 4$ , to neglect issues  related to the logarithmic  corrections to the  soft factors.} dimensional gravitational coupling constant ,   $\epsilon_{MN}=\epsilon_N\,\bar{\epsilon}_M$, $M,N=0\dots d$,  denotes the polarization of the soft graviton decomposed in its left and right components. In these notes,  we shall be interested in the leading and subleading  soft operators whose explicit  forms are
\begin{eqnarray}
\hat{S}^{(-1)}=   \epsilon_{MN}\sum_{i=1}^n\frac{p_i^M\,p_i^N}{p_i\cdot q}\qquad,\qquad\hat{S}^{(0)}=   \epsilon_{MN}\,\sum_{i=1}^n  \frac{q_Pp_i^M\, J_i^{NP}}{p_i\cdot q} \label{2.2}
\end{eqnarray} 
with $J_i^{MN}$  the total angular momentum operator acting on the polarization tensors of finite energy states inside $M_n$. It is given by the sum of orbital and spin angular momentum 
\begin{eqnarray}
J_i^{MN}= L_i^{MN}+S_i^{MN}~~ ;~~ L_i^{MN}=p_i^M\frac{\partial}{\partial p_{iN}}-p_i^N\frac{\partial}{\partial p_{iM}}~~.
\label{2.3}
\end{eqnarray}
The spin angular momentum operator $S_i^{MN}$ takes different representations depending on what finite energy state it acts upon. 
The leading and the subleading soft operators $\hat{S}^{(-1)}$ and $\hat{S}^{(0)}$ are universal while  the sub-subleading operator $\hat{S}^{(1)}$, whose explicit form is given in \cite{1706.00759}, is theory dependent. In Heterotic and Bosonic string theory, for example,  it is  modified by terms  proportional to $\alpha'$,  the string slope \cite{1406.5155, 1610.03481} due to the presence of couplings $\phi R^2$.

The universality of the first two terms in equation \eqref{1soft} allows us to apply the theorem to an arbitrary  $d+1$ dimensional theory  of gravity and  for amplitudes with generic finite energy states  even though we shall only consider finite energy gravitons . We compactify the theory  on a  circle of radius $R_d$ parametrized by the coordinate $0\leq z\leq 2\pi R_d$. The dimensional reduction determines three different soft-factorization properties,  namely, those of the $d$-dimensional graviton, scalar $\phi$  and  vector field $A_\mu$. These states are those emerging from the dimensional reduction of the $d+1$ dimensional graviton polarization tensor according to the identifications
\begin{eqnarray}
\epsilon_{\mu\nu}(p^\mu) &=& \frac{\kappa_{d}}{\kappa_{d+1}} \left( \varepsilon_{\mu\nu}(p^\mu)+\frac{2\alpha }{\sqrt{2}}\hat{\phi}(p^\mu)\,\eta^\perp_{\mu\nu} \right)+  O(\kappa_{d}^2)\nonumber\\[.3cm]
\epsilon_{\mu z}(p^\mu)&=& \frac{\kappa_{d}}{\sqrt{2}\kappa_{d+1}} \varepsilon_\mu(p^\mu)+O(\kappa_{d}^2)~~;~~\epsilon_{zz}(p^\mu)= 2\beta \frac{\kappa_{d}}{\sqrt{2}\kappa_{d+1}}\hat{\phi}(p^\mu) +O(\kappa_{d}^2)~~.
\label{2.14}
\end{eqnarray}
Here,  $\varepsilon_\mu$, $\mu=0\dots d-1$,  denotes the polarization of the vector field in $d$ dimension and (on-shell)
\begin{eqnarray}
\eta_{\mu\nu}^\perp\equiv\eta_{\mu\nu} -p_\mu \bar{p}_\nu -p_\nu\bar{p}_\mu\quad;\quad p\cdot \bar{p}=1\quad ;\quad p^\mu\eta_{\mu\nu}^\perp(p)=0\qquad;\qquad \bar p^2=0~~.
\end{eqnarray} 
The constants $\alpha$ and $\beta$ are chosen to be
\begin{eqnarray}
\alpha^2= \frac{1}{2(d-2)(d-1)}\qquad, \qquad \beta =-(d-2)\alpha~~.\label{2.9}
\end{eqnarray}
These values guarantee that the compactification of the Einstein-Hilbert action gives the $d$-dimensional action in the Einstein frame \cite{cho&zho}. The compactification  also produces an infinite tower of massive spin-$2$ states  which are the Fourier modes of the $d$-dimensional fields charged with respect to the gauge field and with $KK$ charges given by $e_p=\frac{n}{R_d}$  where $p\in \mathbb{Z}$.

In the following,  we aim to get the gyromagnetic factors of the $KK$-massive spin-$2$ states and, therefore, we discuss only the soft theorem of the vector field. The scalar soft theorem  is analysed in detail in \cite{1911.05099}. The compactification of equation \eqref{1soft}  gives the following soft behaviour, valid up to subleading order,  of the vector \cite{1911.05099}
\begin{eqnarray}
{\cal M}^A_{n+1}
&=&\sqrt{2}\,\kappa_d \sum_{i=1}^n\varepsilon_\mu  \left[\frac{e_{p_i} p_i^\mu}{p_i\,q}+\frac{e_{p_i}\, \,q_\nu \left(2\,L_i^{\mu\nu}+S_i^{\mu\nu}\right)}{2p_i\, q} +\frac{ q_\nu p_i^\sigma (\Sigma_{\sigma \rho})^{\mu\nu}S_i^{z\rho }}{2p_i\, q} 
\right]{\cal M}_n~~.
\label{2.39}
\end{eqnarray}
Here, $e_{p_i}$ are the KK-charges of the  finite energy states, $L_i$ and $S_i$ are the $d$-dimensional angular and spin operators, respectively and $(\Sigma_{\sigma \rho})_{\mu\nu}\equiv\eta_{\sigma \mu}\,\eta_{\rho\nu}-\eta_{\sigma \nu}\,\eta_{\rho \mu}$. $S_i^{\rho z}$ is the $d+1$  dimensional  spin operator with one index extended along the compact direction. Its action on the massive fields annihilates them because it gives polarizations which are eaten by the $d$-dimensional massive spin-$2$ particles .

The soft-theorem for vector field in interaction with such KK-states can also be obtained from explicit amplitude calculations starting from  the Fierzi-Pauli action minimally coupled with an Abelian $U(1)$ gauge field. This is the action describing the massive spin-$2$ states which can be obtained, for example, from the compactification of the Einstein-Hilbert  action keeping all the massive modes. The explicit expression of such action is given in\cite{cho&zho}. Here we only give the interaction vertex among a vector, with momentum $q$ and polarization $\varepsilon^\mu$, and two massive particles,  having momenta and polarizations $(k_2, \phi_{\mu \nu})$ and $(k_3,{\phi^*}_{\rho \sigma})$. This vertex is necessary to compute the Feynmann diagrams contributing to the tree-level scattering of a vector in interaction with an arbitrary number of HS-particles (see Fig.\ref{fig1}):
  \begin{eqnarray}
&&V_{\tau ;\rho \sigma;\mu \nu}(q,k_3,k_2)=\frac{i}{2} {e_q}\, \left[V^{(0)}_{\tau ;\rho \sigma;\mu \nu}(q,k_3,k_2)+g\,V^{(g)}_{\tau ;\rho \sigma;\mu \nu}(q,k_3,k_2)\right]
\end{eqnarray}
with\footnote{Here, we have introduced the notation $\{a,\,b\}=\frac{1}{2} (a\,b+b\,a)$.}
\begin{eqnarray}
&&V^{(0)}_{\tau ;\rho \sigma;\mu \nu}=\frac{1}{2}(\eta_{\rho \mu}\eta_{\sigma\nu} +\eta_{\rho\nu}\eta_{\sigma \mu} -2 \eta_{\rho \sigma}\eta_{\nu\mu}) (k_2-k_3)_\tau+ \eta_{\tau \{\rho}\eta_{\mu \nu}(k_2-k_3)_{\sigma\}}\nonumber\\
&&+ \eta_{\tau \{\mu} \eta_{\rho \sigma} (k_2-k_3)_{\nu\}} - \eta_{\tau \{\rho} (\eta_{\sigma\}\nu} k_{2\mu}+ \eta_{\sigma\}\mu} k_{2\nu}) + \eta_{\tau \{\mu} (k_{3 \rho}\eta_{\nu\}\sigma} +\eta_{\nu\}\rho } k_{3 \sigma})
\end{eqnarray}
and 
\begin{eqnarray}
&&V^{(g)}_{\tau ;\rho \sigma;\mu \nu}=
\eta_{\tau \{\mu} (q_{\sigma} \eta_{\nu\}\rho}+ q_{\rho} \eta_{\nu\}\sigma}) +q_{\rho} \eta_{\sigma\mu})-\eta_{\tau \{\sigma}(q_{\mu} \eta_{\rho\} \nu}+ q_{\nu} \eta_{\rho\}\mu})~~.
\label{4.65b}
\end{eqnarray}
The vertex contains an arbitrary gyromagnetic ratio $g$ which takes into account the ambiguity, due to the non-commutativity of the covariant derivatives, in applying the minimum coupling substitution \cite{Deser&Waldron}.  The on-shell amplitude 
gets contribution from the two diagrams  shown  in Fig.\ref{fig1}.  The  exchange diagram and  the  contact term $N_{n+1}^\mu(q,\,p_i,\,  p_{n})$ are given by:
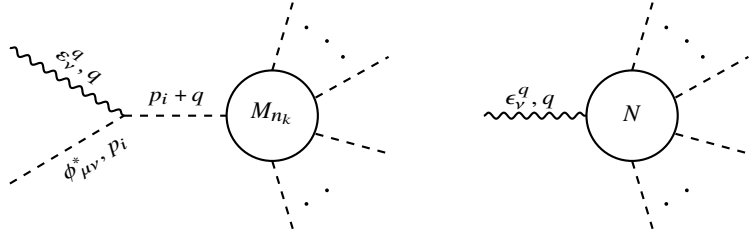
\begin{figure}
	\begin{center}
		\begin{tikzpicture}[scale=.30]
		\draw [thick, dashed]  (8.5,2) -- (9.4,5); 
		\draw [thick, dashed]  (10.4,0.8) -- (13.6,2.6); 
		\draw [thick, dashed]  (10.4,-0.8) -- (13.7,-1.7); 
		\draw [thick, dashed]  (8.5,-2) -- (9.4,-5); 
		\draw [thick, dashed] 
		(2,0)-- (6.5,0)  node[midway, sloped,above] {\scriptsize $\hspace*{.01in} p_i+q$};; 
		\draw [thick,decorate,decoration={snake,amplitude=.4mm,segment length=2mm,post length=0mm}] (-3,3) -- (2,0) node[midway, sloped,above] { \scriptsize $\hspace*{.01in}\varepsilon^q_\nu, q$};
		\draw[thick, dashed]  (-3,-3) -- (2,0) node[midway, sloped,below] { \scriptsize $\hspace*{.2in}{\phi^*}_{\!\!\!\mu\nu}, p_{i}$};
		\draw (8.5,.1) node { \footnotesize $M_{n_k}$};
		\draw [thick] (8.5,0) circle (2cm);
		\begin{scope}[shift={(-3,0)}] 
		\filldraw [ thick] (13.0,3.9) circle (1pt);
		\filldraw [ thick] (13.9,3.3) circle (1pt);
		\filldraw [ thick] (14.6,2.6) circle (1pt);
		\filldraw [ thick] (13.0,-3.9) circle (1pt);
		\filldraw [ thick] (13.9,-3.3) circle (1pt);
		\end{scope}
		\end{tikzpicture}
		\hspace*{1cm}
		\begin{tikzpicture}[scale=.30]
		\draw [thick, dashed]  (8.5,2) -- (9.4,5); 
		\draw [thick, dashed]  (10.4,0.8) -- (13.6,2.6); 
		\draw [thick, dashed]  (10.4,-0.8) -- (13.7,-1.7); 
		\draw [thick, dashed]  (8.5,-2) -- (9.4,-5); 
		\draw [thick,decorate,decoration={snake,amplitude=.4mm,segment length=2mm,post length=0mm}]        (2,0)   -- (6.5,0); 
		\draw (4,.8) node { \scriptsize $\epsilon_\nu^q, q$};
		\draw (8.5,.1) node { \footnotesize $N$};
		\draw [thick] (8.5,0) circle (2cm);
		\begin{scope}[shift={(-3,0)}] 
		\filldraw [ thick] (13.0,3.9) circle (1pt);
		\filldraw [ thick] (13.9,3.3) circle (1pt);
		\filldraw [ thick] (14.6,2.6) circle (1pt);
		\filldraw [ thick] (13.0,-3.9) circle (1pt);
		\filldraw [ thick] (13.9,-3.3) circle (1pt);
		\end{scope}
		\end{tikzpicture}
		\caption{Diagrams collecting the interactions of a gauge field with $n$ KK-particles.}  
		\label{fig1}  
	\end{center}
\end{figure}
 \begin{eqnarray}
M_{n+1}&\equiv& \varepsilon^\tau (M_{n+1})_{\tau}(q,\,p_1\dots, p_{n}) \nonumber\\
&=&\varepsilon^\tau \, {\phi^*}^{\rho\sigma}\sum_{i=1}^{n}\left[V_{\tau ;\rho\sigma;\alpha\beta}(q,p_i,-p_ i-q)D^{\alpha\beta\mu\nu}(-p_i-q) M_{\mu\nu}(p_1,\dots p_i+q\dots,p_{n})\right]\nonumber\\[.1cm]
&&+\ \varepsilon_{\mu}N_{n+1}^\mu(q,\,p_i,\,  p_{n}) 
\end{eqnarray}
with $D^{\alpha\beta\mu\nu}$  being the propagator of the massive spin two field whose explicit expression is, for example, given in \cite{Deser&Waldron}. By imposing the current conservation condition \cite{1406.6987}
\begin{eqnarray}
(p_i+q)^\mu \,M_{\mu\nu}(p_1,\dots p_i+q,\dots p_n)=(p_i+q)^\nu \,M_{\mu\nu}(p_1,\dots p_i+q,\dots p_n)=0,
\end{eqnarray}
and taking the external particles on-shell, it is possible to determine the contact contribution $N_{n+1}^\mu(q,\,p_i,\,  p_{n})$ up to ${\cal O}(q)$ in the expansion of the vector momentum,  obtaining
\begin{eqnarray}
M_{n+1}=\varepsilon_{\mu} \sum_{i=1}^{n} {e}_{q_i}\,\Bigg[\frac{p_{i}^\mu}{p_i q} +\frac{q_\rho  }{2 p_iq}( L^{\mu\rho}_i+gS_i^{\mu\rho}) +\frac{ q_\rho}{  2p_iq}
L_i^{\mu\rho}\Bigg]M_n(p_i)+O(q)~~.
\end{eqnarray}
This equation is consistent with the result given in equation \eqref{2.39},  specialized to finite energy massive spin-$2$ particles, (in which case, the last term in the right-hand side of \eqref{2.39} vanishes) provided we take the gyromagnetic ratio to be $g=1$.

\section{Gyromagnetic Ratios from Superstring Theories}
\label{gyrofromstring}
String theories with their infinite towers of massive high spin excitations provide a quantum description of high spin theories with an infinite number of massive fields.
They are formulated in higher-space time dimensions where the conformal anomalies cancel and the compactification of the extra dimensions is required to connect the theory to the real world. The compactification breaks the higher dimensional Lorentz group and creates new gauge fields arising from higher dimensional tensors transforming as vectors under the unbroken  Lorentz  group. In the following, we shall consider the toroidal compactification of Heterotic, Type II and Bosonic string theories and focus our analysis on the $U(1)$ gauge fields emerging from the dimensional reduction of the graviton and Kalb-Ramond state.  KK-modes of the HS-string states will be charged with respect to  these Abelian fields and  we aim to derive an expression of their gyromagnetic ratios. 

The compact manifold that we introduce to reduce the space dimensions is the $D$-dimensional  torus viewed as the quotient space $T^D=\frac{\mathbb{R}^D}{2\pi \Lambda_D}$ with  $\Lambda_D$ a $D$ dimensional lattice. A complete set of linearly independent vectors on the lattice is denoted by  $\textbf{e}_i\equiv\{e_i^a\}$  which also define the  torus metric
\begin{eqnarray} 
g_{ij}= e_i^{a}\, \, e_j^b\delta_{ab}\qquad;\qquad i,\,j=1,\,\dots D~~;~~\,a,\,b=1,\,\dots D~~.\label{7n}
\end{eqnarray} 
The dual lattice  is introduced through the dual base vectors $\textbf{e}^{*i}=\{e^{*i}_a\}$ satisfying  the relations
\begin{eqnarray}
e^{*i}_b\,e^a_i=\delta^a_b~~;~~ e_i^a\,e^{*j}_a=\delta^j_i~~;~~\sum_{a=1}^D e^{*j}_a\,e^{*i}_a=g^{ij}
\end{eqnarray}
We  consider the propagation of  strings in a generic non-constant  background with metric  and anti-symmetric field expressed by the compactification ansatz
\begin{eqnarray}
G_{MN} =\Bigg( 
\begin{array}{ccc}
\eta_{\mu\nu}+g_{kl} A_\mu^k(X^\mu)\,A_\nu^l(X^\mu)&& A_{\mu j}(X^\mu)\,\\ 
A_{i\nu}(X^\mu)\,&& g_{ij}
\end{array}\Bigg) ~~;~~B_{MN}=\Bigg(
\begin{array}{cc}
0&B_{\mu j}(X^\mu)\\
{B_{i\nu }(X^\mu)} &B_{ij} \end{array}
\Bigg) ~~;~~\label{3.5.4}
\end{eqnarray}
with $\mu, \nu=0,\dots d-D-1$, $d$ being the critical dimensions of the string theory and $g_{ij}$ and $B_{ij}$ constant moduli of the torus $T^D$. $A_{\mu i}(X^\nu)$ and $B_{\mu i}(X^\nu)$ are the lower dimensional $U(1)$ gauge fields which are assumed to have constant field strengths $F^A_{\mu\nu;\,i}$ and $F^B_{\mu\nu;\,i}$, respectively in the forthcoming discussion.
 
The  Bosonic string coordinates $X^i(\uptau,\sigma)$, $i=1,\dots D$, extend along the compact directions of the manifold where they satisfy the periodic identifications
\begin{eqnarray}
 X^i(\uptau,\sigma +\ell) = X^i(\uptau,\sigma) + 2\pi\sqrt{\alpha'}\ n^i\qquad,\quad n^i\in \mathbb Z~~.
 \end{eqnarray}
The $\ell$ can be either $\pi$ or $2\pi$ depending on the adopted conventions and $\alpha'$ is the string slope.
 
 In this compact background, we consider the $(1,1)$ supersymmetric sigma model in ten dimensions.  This corresponds to Type II superstring theory. It is described by the world-sheet action:
 \begin{eqnarray}
S&=&\frac{1}{4\pi\alpha'} \int d^2\sigma \bigg[4G_{MN}\partial_+X^M\partial_-X^N+4B_{MN}\partial_+X^M\partial_-X^N +2iG_{MN}\psi_+^M\tilde{\nabla}_-\psi_+^N \nonumber\\
&&\hspace*{.87in}+\ 2iG_{MN} \psi_-^N\tilde{\nabla}_+\psi_-^M+\frac{1}{2}\tilde{R}_{MNPQ}\psi_+^M\psi_+^N\psi_-^P\psi_-^Q\bigg]~~.\label{4.3.45}
\end{eqnarray}
 Here,  $\psi_\pm(\uptau, \sigma)$ are the world-sheet supersymmetric partners of the bosonic coordinates and the covariant derivatives 
   $\tilde\nabla_\pm$ are defined by
\begin{eqnarray}
\tilde\nabla_\pm \psi^M_\mp=\partial_\pm \psi_\mp^M+\tilde\Gamma^M_{\pm\,PQ}\psi_\mp^P\ \partial_\pm X^Q\quad,\qquad \tilde\Gamma^M_{\pm PQ}=\Gamma^M_{\;\;PQ}\pm\frac{1}{2}H^M_{\;\;PQ}
\end{eqnarray}
The $\tilde{\Gamma}^P_{\pm MN}$ are the connections with a totally antisymmetric torsion. The $\tilde R_{MNPQ}$ are given by
\begin{eqnarray}
\tilde R_{MNPQ}=R_{MNPQ}+\frac{1}{2} \nabla_PH_{MNQ}-\frac{1}{2} \nabla_QH_{MNP}+\frac{1}{4} H_{MRP}H^R_{\;\;QN}-\frac{1}{4} H_{MRQ}H^R_{\;\;PN}~~.
\end{eqnarray}
The information on the gyromagnetic ratios of the massive string states charged with respect to the emerging Abelian gauge fields is encoded in the part of string Hamiltonian which is linear in the fields $A_{\mu i}$ and $B_{\mu i}$. For convenience, we separate such contributions into two classes of interactions.  Those giving  the interaction of one gauge field with the world-sheet fields:
\begin{eqnarray}
H_1&=& \frac{1}{2\pi\alpha'}\int_0^\ell d\sigma \biggl[A_{\mu i}\Bigl\{-(2\pi\alpha')^2\Pi^\mu\Pi^i-(2\pi\alpha')\Pi^\mu B^i_{~\;j}\partial_\sigma X^j+\partial_\sigma X^i \partial_\sigma X^\mu\nonumber\\
&&\hspace*{.5in}+\ \frac{i}{2}\Bigl(  \psi^\mu_+\partial_\sigma\psi^i_+ -\psi^\mu_-\partial_\sigma\psi^i_-+ \psi^i_+\partial_\sigma\psi^\mu_+ -\psi^i_-\partial_\sigma\psi^\mu_-\Bigl) \Bigl\}\nonumber\\
&&\hspace*{.5in}+\ B_{\mu i}\Bigl\{  (2\pi\alpha') \Pi^\mu \partial_\sigma X^i+(2\pi\alpha')\Pi^i\partial_\sigma X^\mu  +g^{ij}B_{jk}\partial_\sigma X^k \partial_\sigma  X^\mu\Bigl\}\biggl]\label{4.5.113ed}
\end{eqnarray}
and the terms describing the interactions among the world-sheet  and the field strengths of the background gauge fields:
\begin{eqnarray}
H_2&=&- \frac{i}{4}\int_0^\ell d\sigma \biggl[F^A_{\mu\nu;\,i}\Bigl\{\Pi^\mu\Psi_+^{\nu i}-\Pi^i\Psi_+^{\mu \nu}-\frac{1}{2\pi\alpha'}\Bigl( B^i_{\;\;\;j}\partial_\sigma X^j\Psi_+^{\mu\nu}+\partial_\sigma X^\mu\Psi_-^{\nu i} -\partial_\sigma X^i\Psi_-^{\mu \nu}\Bigl) \Bigl\}\nonumber\\
&&+F^B_{\mu\nu; i}\Bigl\{ -2\Pi^\mu \Psi_-^{\nu i}-\Pi^i\Psi_-^{\mu\nu}  +\frac{1}{2\pi\alpha'} \Bigl(-B^i_{\;\;j}\partial_\sigma X^j\Psi_-^{\mu\nu}+2\partial_\sigma X^\mu\Psi_+^{\nu i}+\partial_\sigma X^i\Psi_+^{\mu \nu} \Bigl)\Bigl\}\biggl]
\label{59}
\end{eqnarray}
where, we have defined $\Psi_\pm^{MN}=\psi_+^M\psi_+^N\pm \psi_-^M\psi_-^N$. The expectation value of the interacting Hamiltonian between two generic string states of mass $m$ is given by 
\begin{eqnarray}
\langle \Phi|{\cal H}_I|\Phi\rangle= \langle \Phi|\frac{l}{2\pi\alpha' m }H _1|\Phi\rangle+\langle \Phi|\frac{l}{2\pi\alpha' m }  H_2|\Phi\rangle
\end{eqnarray}  
where the normalization factor $\frac{l}{2\pi\alpha' m }$  is introduced to  relate the string Hamiltonian to that of charged massive point particles \cite{9405117}.  $\Phi$ is a generic  physical state of the closed superstring spectrum. By using the expression of the string fields given in terms of creation an annihilation oscillators and introducing the constant field strength of the  $U(1)$-gauge fields,   one gets
\begin{eqnarray}
\langle\Phi|{\cal H}_I|\Phi\rangle
&=&-\frac{1}{2 m} F^A_{\mu\nu;\,i} \Bigl\langle\Phi\Bigl|\frac{1}{2}L^{\mu\nu} \bigl( p_R^i+p^i_L\bigl)\ +\ p^i_RS_L^{\mu\nu}  \:+\  p^i_LS_R^{\mu\nu}\Bigl|\Phi\Bigl\rangle\nonumber\\
&&-\frac{1}{2m} F^B_{\mu\nu;\,i} \langle\Phi |\frac{1}{2}L^{\mu\nu}  \bigl( p_R^i-p^i_L\bigl)\ -\ S_R^{\mu\nu} p^i_L \:+\ S_L^{\mu\nu} p^i_R |\Phi \rangle~~.\label{hib}
\end{eqnarray}
In the above expression,  $p_R$ and $p_L$ are the compact momenta defined in \cite{9401139},  $S_L$ and $S_R$ are the usual left and right spin operators that in superstring theories get contributions  from the bosonic and fermionic world-sheet oscillators \cite{GSWI}. 
{Here, equation \eqref{hib} has been derived in Type II theory. However, in 
\cite{2102.13180}, they are also studied in the Bosonic and Heterotic sigma-models in arbitrary compact toroidal backgrounds. It  turns out  that  expression \eqref{hib} is universal being invariant in form for all the cases considered, the only  difference is in the definition of the spin operators which are theory dependent.  

Gyromagnetic factors, as shown in\cite{2102.13180}, can also be extracted from $3$-point amplitudes involving a graviton vertex in interaction with arbitrary string states. The compactification of these amplitudes produces the $U(1)$ gauge fields with their minimal and non-minimal couplings  described by the  effective action given in equation \eqref{APhiPhi}.  In this approach the gyromagnetic factors are entirely encoded in the   minimal couplings of the  gravity before the compactification. Thus these are unique and translate into highly constrained electromagnetic couplings in the lower dimensional theory. The expressions obtained from the amplitude  and the Hamiltonian formalism  are  consistent  and give the following  electromagnetic coupling of the HS-states
 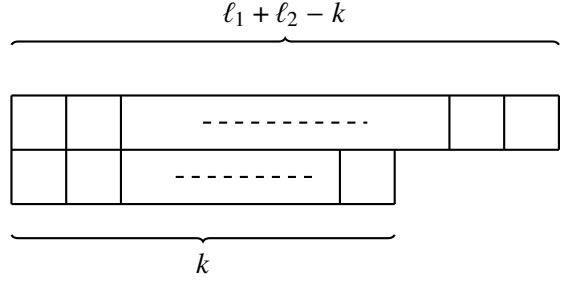
\begin{figure}[t]
\begin{center}\hspace*{-.6in}
\begin{tikzpicture}[scale=.36]
\draw [thick]  (-10,2) -- (10,2);
\draw [thick]  (-10,0) -- (10,0);
\draw [thick]  (-10,2) -- (-10,0);
\draw [thick]  (10,2) -- (10,0);
\draw [thick]  (-8,2) -- (-8,-2);
\draw [thick]  (8,2) -- (8,0);
\draw [thick]  (-6,2) -- (-6,-2);
\draw [thick]  (-10,-2) -- (4,-2);
\draw [thick]  (-10,0) -- (-10,-2);
\draw [thick]  (4,0) -- (4,-2);
\draw [thick]  (2,0) -- (2,-2);

\draw [thick]  (6,2) -- (6,0);
\draw [thick,dashed]  (-3,1) -- (3,1);
\draw [thick,dashed]  (-4,-1) -- (1,-1);

\draw[decorate, decoration={brace, mirror}, yshift=5ex,thick]  (10,3) -- node[above=0.6ex] {$\ell_1+\ell_2-k$}  (-10,3);
\draw[decorate, decoration={brace,mirror}, yshift=5ex,thick]  (-10,-4) -- node[below=0.6ex] {$k$}  (4,-4);

\end{tikzpicture}
\end{center}
\caption{Young diagram with two rows}
\label{Younghook}
\end{figure}
 \begin{equation}\label{symm_id}
   g^a_{A;B}=\frac{1}{p_L^a\pm p_R^a } \Big\langle\Phi\Big|p_R^a S^{\mu\nu}_L\pm p_L^a S^{\mu\nu}_R \Big|\Phi \Big\rangle~~.
\end{equation}
 This expression was originally obtained for Bosonic string in \cite{9405117} and extended to Heterotic and Type II string theories in \cite{2102.13180}. 
 
 Eq. \eqref{symm_id} is still a formal expression,  to  extract the explicit value of the gyromagnetic factors. one needs to consider explicit examples. For $p_R=\pm p_L$, which corresponds to vanishing Kaluza Klein ($p_L=-p_R$) or Winding charges ($p_L=p_R$),   the relevant interactions depend on the combination $S=S_L+S_R$ which allows us to read off the gyromagnetic ratio for arbitrary elements of the spectrum regardless of the Young Tableau representation of the string states. The gyromagnetic factor, in this case, is $g=1$. This includes the result  obtained in Sec.\ref{soft} for the KK spin-$2$ particles via the soft theorem and generalizes 
  previous results obtained in different contexts\cite{HIOY:84,9801072}.

 The mixed symmetry case is more instructive and is relevant  when the non-compact directions are larger than four. We focus here on the example of two row Young Tableaux which appear in the first Regge trajectory of the closed Bosonic string.
 In this case, one starts from the product of two totally symmetric representations of spins $\ell_R$ and $\ell_L$ with $\ell_R\geq \ell_L$, associated with the first Regge trajectory of the open string and project onto the irreducible components associated to the tableaux $\{\ell_R+\ell_L-k,k\}$, $k\leq \ell_L$ (see Fig. \ref{Younghook}). The resulting HS-state, in the compact representation of the physical states discussed in  Section \ref{notation}, is   
\begin{align}
    \phi_{\ell_L+\ell_R-k,k}={\mathcal{N}}_{\ell_L+\ell_R-k,k}(u\cdot w)^{\ell_L-k} ({u}\cdot \bar{w} )^{\ell_R-k}\left(u\cdot w \, \bar{u}\cdot \bar{w}- {u}\cdot \bar{w}\,  \bar{u}\cdot w\right)^k\,.
\end{align}
The gyromagnetic ratios follow  from the following identity which is derived in \cite{2102.13180} 
\begin{align}
    \Big\langle\Phi\Big|p_R S^{\mu\nu}_L+p_L S^{\mu\nu}_R\Big|\Phi\Big\rangle=\big\langle\Phi\big|\alpha_1 S_1^{\mu\nu}+a_2S_2^{\mu\nu}\big|\Phi\big\rangle_u\,,
\end{align}
where on the left hand side we have the closed string-correlator and on the right-hand side we used the inner-products among Young Tableaux. Here, $S_1$ and $S_2$ are defined in equation \eqref{5} and
\begin{eqnarray}
    \alpha_1=\frac{(\ell_R-k) p_R+(\ell_L-k) p_L}{\ell_R+\ell_L-2}\,\qquad,\qquad
    \alpha_2=\frac{(\ell_R-k) p_L+(\ell_L-k) p_R}{\ell_R+\ell_L-2}\,.\label{genalph}
\end{eqnarray}
We can then read off the gyromagnetic ratios for the gauge field $A^a_\mu$
\begin{align}
    g_1^{(a)}&=\frac{2}{p_L^a+p_R^a}\frac{(\ell_R-k) p_L^a+(\ell_L-k) p_R^a}{\ell_R+\ell_L-2k}\,,\\
    g_2^{(a)}&=\frac{2}{p_L^a+p_R^a}\frac{(\ell_R-k) p_R^a+(\ell_L-k) p_L^a}{\ell_R+\ell_L-2k}\,,
\end{align}
as well as for the gauge field $B^a_\mu$
\begin{align}
    g_1^{(a)}&=\frac{2}{p_L^a-p_R^a}\frac{(\ell_R-k) p_L^a-(\ell_L-k) p_R^a}{\ell_R+\ell_L-2k}\,,\\
    g_2^{(a)}&=\frac{2}{p_L^a-p_R^a}\frac{-(\ell_R-k) p_R^a+(\ell_L-k) p_L^a}{\ell_R+\ell_L-2k}\,.
\end{align}
Similar expressions can be obtained for any mixed-symmetry representation in subleading Regge trajectories.
 \section{Conclusions}  
\label{Conclusion}
 In these notes, we have discussed interesting relationships emerging from the compactification of soft theorems and the formulation of string theories in arbitrary compact toroidal backgrounds determining the electromagnetic couplings of massive spinning particles carrying KK and winding charges.  A universal expression for the gyromagnetic factors has been derived holding for all the string states minimally coupled to Abelian gauge fields arising from the dimensional reduction of the graviton and Kalb-Ramond state.  The same results have also been obtained from the compactification of three-point  string amplitudes with a  massless state of the NS-NS-sector in interaction with arbitrary massive string states, evaluated in Bosonic, Heterotic and Type II string theories. These amplitudes have been extensively discussed in \cite{1911.05099}.

We have followed a top-down approach.  From this perspective, the universality of the gyromagnetic ratios, for the states carrying only KK-charges,   follows both from the universality of gravitational soft theorems and from their minimal couplings with the higher dimensional gravitons from which they originate via compactification. However, similar considerations cannot be extended to generic string states, of the models discussed above, carrying also arbitrary winding charges.  For these, the origin of the universality of their couplings with the Abelian gauge fields, arising from compactification,  is still obscure.   
 
It would be interesting to get the same results directly from the lower dimensional higher spin theories where, by analogy with the case $g=2$,  unitarity based arguments applied to the high energy limit of Compton scattering amplitudes could explain the general expression found for electromagnetic couplings of generic compact HS-states. Furthermore, it would also  be compelling  to explore  the  gyromagnetic couplings of massive spinning particles  in other related set-ups such as orbifold, flux compactification or in ADS backgrounds where these results could  be mapped in some property of the  dual  CFT  correlators, along  similar patterns analysed in \cite{Camanho:2014apa}.

\bigskip

{\bf Acknowledgement}: We would like to thank M. Taronna for the relevant contributions given to this project and summarized in these notes.

\end{document}